# HAC Analysis to Explore Clusters within Chronic Comorbid Inpatient Visits

Rasika Karkare

Department of Mathematical Sciences,

Worcester Polytechnic Institute,

Worcester, MA, USA

1. Abstract:

Multimorbidities are associated with significant burden on the healthcare system and the lack of accurate and pertinent statistical exploratory techniques have often limited their analysis. Here we employ exploratory hierarchal agglomerative clustering (HAC) of multimorbidities in the inpatient population in the state of Ohio. The examination exposed the presence of ten discrete, clinically pertinent groups of multimorbidities within the Ohio inpatient population. This method offers an assessable empirical exploration of the multimorbidities present in a specific geographic populace.

2. Introduction:

Patients with more than one chronic conditions are referred to as multimorbid patients. The incidence of such multimorbid patients has increased due to the rise in the aging population in the US leading to significant impact on the overall health care outcomes. Numerous mechanisms may lie beneath these multimorbidities, including direct relationship, associated risk features, heterogeneity, and individual differences [1,2]. There has been an augmented acknowledgment of its bearing and the significance of improving consequences for such multimorbid patients[3–5]. The lack of proper and statistical exploratory techniques have often limited the analysis of such chronic multimorbidities.

Clustering, is an unsupervised data mining algorithm that groups similar units into similar clusters, thus partitioning dissimilar objects into other clusters[6–10]. Hierarchical agglomerative clustering is a repeated subdivision of a dataset into clusters at an progressively finer granularity[11]. HAC has previously been employed to find clusters in microarray data[12], employed in nursing research[10], and to find molecular genetic markers[9]. In the current study, we use unsupervised clustering exploration to discover the cohorts of cohorts present in the Ohio patient population. The described clustering method classifies units of multi-morbidities and presents prospects for improved supervision of multimorbid patients.

3. Methods:

3.1 Study Population

We used open access de-identified aggregate data provided by the Ohio Department of State Health Services for this research. Several exclusions including pregnant females, cancer, patients in hospice or long term care were excluded from the analysis. Further, patients only within the ages of 40 to 80 were identified for this multimorbidity analysis. Identification of conditions within unit associates were founded on an in/outpatient data matched to *International Classification of Diseases, Tenth Revision (ICD-10)* diagnosis codes in 2015[13]

3.2 Exploratory Data Analysis and Feature Generation

Apache Hadoop database was used to store, query and extract the dataset extracted from the online resource. One step hot encoding was used for dichotomous target features for selected disorders. Demographic variables were appended to the data set to enable broader interpretation. Data missing at random was imputed by median/mode for continuous/categorical features.

3.3 Statistical Analysis

The R suite of statistical programs was used for quantitative analysis (Version 0.9). Statistical procedures used are described previously in other studies [6,14–16,16–20].

3.4 Clustering Algorithm

An agglomerative hierarchical clustering (AHC) algorithm with a bottom up approach was used to separate clinically appropriate clusters within the study population. The bottom up aproach to AHC initiates with each member starting at an isolated cluster, followed by serial merging of similar members to form similarity clusters until only once cluster remains. After the clustering procedure terminates, subject matter expertise, clinical relevance and study design criterion are used to select a cutoff/threshold which produces the final clusters. The process can be visualized using dendrograms. We used Ward's method along with Gower's distance matrix for similarity calculations as it has shown to be more reliable for mixed data with a preponderance of weighted binary data (like condition related binary variables)[21].

4. Results:

4.1 Clustering Analysis reveals ten broad cluster of multi-morbidity patients in the population

Agglomerative Hierarchal Clustering revealed 10 broad clusters in the population data of 14,444 patients (Figure 1). The average age of patients was 73.0 years and contained 48.7 % males and 58.3 % females.

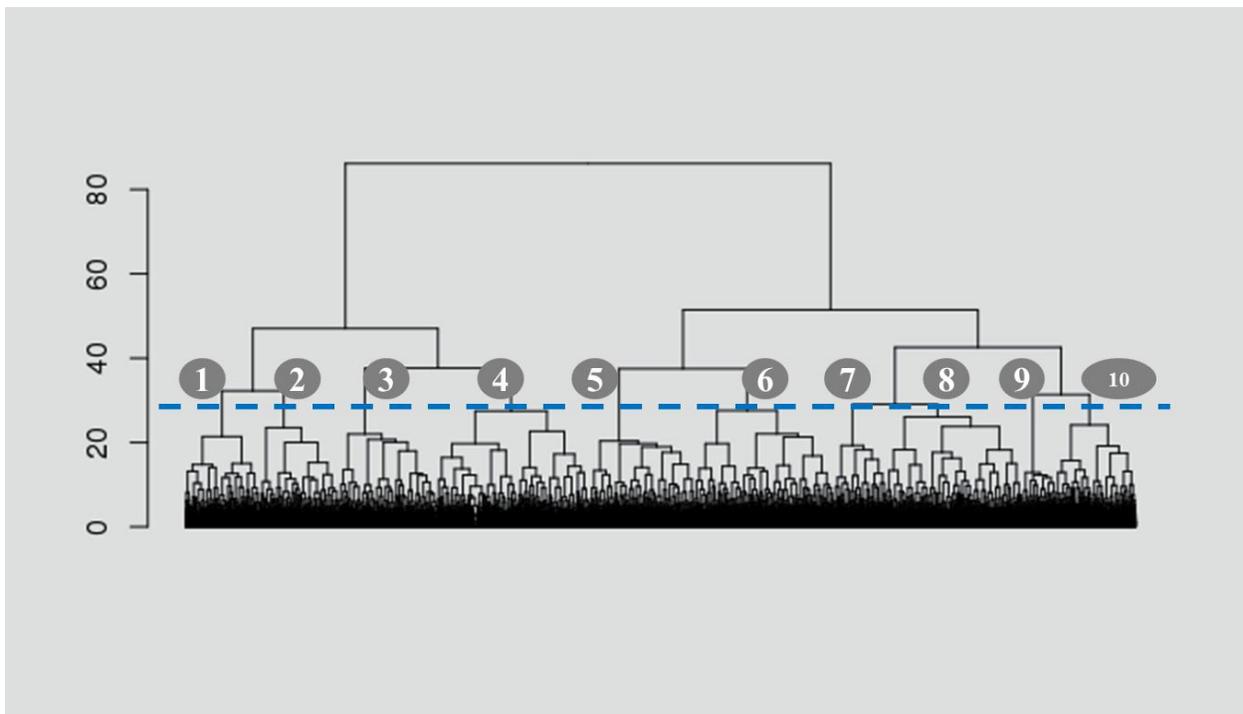

*Figure 1: Results from hierarchical agglomerative clustering reveal nine distinct clusters. Clustering was performed using Ward's method with Gower's distance and a threshold (h=27; shown as a dotted line) was used to isolate 10 clusters.*

Descriptive statistics revealed clinical homogeneity within the clusters (Table 1). The clusters (numbered randomly) were divergent based on the mix of multimorbidities observed and age/gender demographics. Clinically relevant summarization showed the presence of distinct clusters with a high proportion of patients with (Table 2) : metabolic syndromes (cluster 1), cancer (cluster 2), gastro related diseases (cluster 3), hypertensive patients (cluster 4), stomach cancer and related conditions (cluster 5), heart failure and COPD (cluster 6), Obesity related morbidities like lower back pain (cluster 7), Epilepsy (cluster 8), osteoarthritis

(cluster 9) and old age related disorders like Parkinson's (cluster 10). Income, medical utilization, inpatient visits were also calculated but are not shown in the current analysis due to HIPPA restrictions.

| Cluster # | 1 | 2 | 3 | 4 | 5 | 6 | 7 | 8 | 9 | 10 |
|---|---|---|---|---|---|---|---|---|---|---|
| size | 1193 | 2319 | 1600 | 2108 | 1139 | 1494 | 1212 | 2065 | 807 | 507 |
| age (mean) | 68.5 | 62.2 | 75.7 | 76.4 | 77 | 72.9 | 69.9 | 77.5 | 79.2 | 81.1 |
| males (%) | 42.5 | 45.2 | 57.7 | 48.8 | 40.9 | 56.5 | 32.6 | 59.3 | 48.9 | 43 |
| females (%) | 57.5 | 54.8 | 42.3 | 51.2 | 59.1 | 43.5 | 67.4 | 40.7 | 51.1 | 57 |
| Hospital Visits (mean) | 0.73 | 0.94 | 1.14 | 0.98 | 0.68 | 0.89 | 0.94 | 0.91 | 0.82 | 1.07 |
| Diabetes Mellitus (%) | 0 | 0.2 | 0.2 | 0.1 | 0.2 | 0.2 | 0.2 | 0 | 0 | 0.2 |
| Heart Failure (%) | 0.4 | 0.4 | 0.4 | 0.4 | 0.2 | 0.7 | 0.2 | 0.2 | 0.5 | 0.2 |
| COPD (%) | 0.2 | 0.6 | 0.3 | 0.4 | 0.5 | 1.5 | 0.1 | 0.2 | 0.4 | 0.2 |
| Ischemic Heart Disease (%) | 0 | 0.2 | 0 | 0 | 0.1 | 0.4 | 0 | 0.1 | 0 | 0 |
| Otitis Media (%) | 63.4 | 16.4 | 83 | 75.7 | 30.5 | 67.9 | 49.6 | 28.2 | 29.5 | 44.4 |
| Peptic Ulcer Disease (%) | 19.4 | 4.7 | 81.9 | 71.5 | 13.3 | 34.1 | 44.6 | 53.9 | 49.4 | 50.3 |
| Hypertension (%) | 32.7 | 23.8 | 35.4 | 67.3 | 25.1 | 23 | 66.5 | 46.9 | 47.8 | 45.6 |
| Epilepsy (%) | 22 | 6.7 | 77.6 | 67.1 | 28.4 | 14.3 | 57.9 | 78.3 | 33.2 | 62.7 |
| Chronic Thyroid Disorders (%) | 1.3 | 1 | 1 | 1.5 | 1.1 | 0.5 | 2.9 | 1.2 | 0.7 | 1.2 |
| Depression (%) | 3.1 | 2.8 | 2.4 | 6.3 | 2.5 | 1.9 | 4.7 | 2.9 | 3.5 | 4.5 |
| Cholelithiasis Cholecystitis (%) | 95.3 | 51.5 | 99 | 98.3 | 93.4 | 88 | 93 | 95.7 | 93.9 | 97 |
| Iron Deficiency Anemia (%) | 1.8 | 7.7 | 4.6 | 3.8 | 1.2 | 3.3 | 7.3 | 5 | 5.1 | 11.8 |
| Osteoarthritis (%) | 31.3 | 17.4 | 24.9 | 34.3 | 28.4 | 24.2 | 33.9 | 26.1 | 37.2 | 25.6 |
| Rheumatoid Arthritis (%) | 23.7 | 32.9 | 20.9 | 38.5 | 24.5 | 13.4 | 70.1 | 17.5 | 18.3 | 71 |
| Colorectal Cancer (%) | 2.6 | 2.5 | 2.9 | 4.4 | 1.8 | 2.3 | 3.1 | 2.3 | 2.9 | 3.4 |
| Lung Cancer (%) | 10.7 | 6.3 | 23.1 | 15.8 | 12.2 | 10.8 | 11.5 | 9.2 | 16 | 15.2 |
| Hyperlipidemia (%) | 60.9 | 17.9 | 27.9 | 49.8 | 89.5 | 14 | 61.6 | 24.3 | 42.5 | 42.2 |
| Cerebrovascular Disease (%) | 4.9 | 2.5 | 2.6 | 5.6 | 9 | 2.4 | 13.7 | 2.4 | 4.8 | 3.9 |
| Headaches Migrane Othrs (%) | 1.4 | 4.4 | 3.3 | 3.6 | 3.7 | 3.1 | 1.3 | 2.5 | 5 | 2.4 |
| Chronic Renal Failure (%) | 0.8 | 3.8 | 1.8 | 2.6 | 1.6 | 2.2 | 3.2 | 5.2 | 3.8 | 2.2 |
| Kidney Stones (%) | 77.9 | 44.8 | 91.2 | 92.4 | 72.5 | 69.5 | 84.4 | 84.3 | 79.3 | 88.8 |
| Diverticular Disease (%) | 14.6 | 11.1 | 33.4 | 35.1 | 11.4 | 14.9 | 26.3 | 46.4 | 36.6 | 64.5 |
| Low Back Pain (%) | 7.7 | 15 | 4.8 | 8.2 | 4.5 | 4.2 | 28 | 7.1 | 6.8 | 18.1 |
| Nonspec Gastr Dyspepsia (%) | 32.1 | 7.5 | 79.3 | 78.9 | 23.4 | 35.4 | 18.2 | 29.9 | 36.4 | 49.5 |
| Sickle Cell Anemia (%) | 3.3 | 4 | 4.8 | 7 | 4.7 | 3.7 | 6.3 | 3.5 | 5.2 | 3.2 |
| Multiple Sclerosis (%) | 6.1 | 4.6 | 4.4 | 13.4 | 11.4 | 2.2 | 11.8 | 5.5 | 12.5 | 11.2 |

| Condition | | | | | | | | | | |
|---|---|---|---|---|---|---|---|---|---|---|
| Inflammatory Bowel Disease (%) | 56.4 | 39.5 | 22.5 | 60.8 | 48.3 | 14.3 | 72.5 | 39.4 | 40.3 | 44.2 |
| Hemo Congntl Coagulopathies (%) | 63.4 | 44.2 | 50.4 | 68.6 | 54.4 | 31.9 | 75.5 | 50.8 | 52.8 | 76.1 |
| Systemic Lupus Ery (%) | 0.2 | 0.6 | 0.3 | 0.8 | 0 | 0.1 | 0.2 | 0 | 0.4 | 0.2 |
| Prostate Cancer (%) | 0.2 | 1.1 | 0.2 | 0.5 | 0.4 | 0.4 | 0.9 | 0.3 | 0.4 | 0.2 |
| Ovarian Cancer (%) | 2.3 | 5.1 | 0.5 | 1.2 | 1.5 | 1.4 | 2.8 | 1.4 | 2 | 1.4 |
| Endometrial Cancer (%) | 1.4 | 1.1 | 0.2 | 1.2 | 0.4 | 0.6 | 3.3 | 0.6 | 0.4 | 0.8 |
| Cervical Cancer (%) | 5.1 | 2.1 | 7.7 | 7.5 | 6 | 5.8 | 2 | 7 | 8.6 | 9.5 |
| Hodgkin Dis Lymphoma (%) | 0.1 | 1 | 0.1 | 0.3 | 0.4 | 0.1 | 0 | 0.2 | 0.5 | 0.2 |
| Leukemia Myeloma (%) | 0.3 | 0.9 | 0.2 | 0.5 | 0.3 | 0.5 | 0.7 | 0.2 | 0.5 | 0.2 |
| Malignant Melanoma (%) | 0.2 | 0.5 | 0 | 0.2 | 0.3 | 0.1 | 0.1 | 0 | 0.1 | 0.2 |
| Head Neck Cancer (%) | 1.1 | 3 | 1.4 | 1.2 | 1.5 | 1.7 | 1.9 | 1.7 | 1.4 | 1.4 |
| Esophageal Cancer (%) | 1.3 | 2.5 | 2.4 | 2.3 | 1.8 | 3.4 | 0.8 | 1.8 | 3.5 | 1.6 |
| Stomach Cancer (%) | 1.3 | 1.1 | 1.3 | 1 | 3.8 | 1 | 1.3 | 3.1 | 2 | 1 |
| Pancreatic Cancer (%) | 0.3 | 2.1 | 0.4 | 0.6 | 1.1 | 0.4 | 0.3 | 1.2 | 0.4 | 0.2 |
| Pancreatitis (%) | 0.2 | 0.6 | 0.1 | 0.1 | 0 | 0.5 | 0.2 | 0.3 | 0.6 | 0.2 |
| Hepatitis (%) | 2.5 | 6.9 | 2 | 4.7 | 1.8 | 1.5 | 3.6 | 1.4 | 2.5 | 2.8 |
| Peripheral Artery Disease (%) | 2.8 | 5.5 | 1.8 | 1.9 | 2.2 | 4.5 | 4.3 | 1.7 | 1.6 | 2.2 |
| Endometriosis (%) | 15.5 | 8.3 | 40.9 | 86 | 24.1 | 12.4 | 44.8 | 38.9 | 41.6 | 60.2 |
| Ventricular Arrhythmia (%) | 0.1 | 0.5 | 0 | 0 | 0 | 0 | 0.2 | 0 | 0 | 0 |
| Lyme Disease (%) | 2.2 | 1.5 | 12.2 | 8.3 | 2.2 | 3.1 | 5.9 | 10.4 | 5.3 | 3.7 |
| Female Infertility (%) | 0 | 0.1 | 0.1 | 0 | 0.2 | 0.1 | 0.2 | 0 | 0.1 | 0 |
| Menopause (%) | 0.1 | 0.1 | 0 | 0 | 0 | 0 | 0 | 0 | 0 | 0 |
| Glaucoma (%) | 3.7 | 2.8 | 1.5 | 3.7 | 4.7 | 1.7 | 10.1 | 2.7 | 3.5 | 4.1 |
| Low Vision Blindness (%) | 9.3 | 3.8 | 17.6 | 22.2 | 9.2 | 10.9 | 14.4 | 10.7 | 52.3 | 31.2 |
| Cataract (%) | 0.6 | 0.6 | 1.7 | 1.5 | 1.3 | 0.9 | 0.9 | 0.5 | 0.4 | 3.2 |
| Other Cancer (%) | 11.7 | 5.2 | 13.9 | 43.1 | 13.9 | 11.1 | 24.1 | 13.3 | 64.6 | 30.4 |
| Dementia (%) | 3.2 | 11.2 | 4.4 | 6.4 | 7.1 | 10.1 | 4.5 | 7.2 | 7.4 | 4.3 |
| Osteoporosis (%) | 5.5 | 4.7 | 9.5 | 11.5 | 8.6 | 16.9 | 10.6 | 8.2 | 10.7 | 86.8 |
| Obesity (%) | 7.3 | 6.6 | 7 | 16 | 20.8 | 2.9 | 24.8 | 7.1 | 9 | 21.3 |
| Oral Cancer (%) | 87.1 | 11 | 43.7 | 59.3 | 11.8 | 17.3 | 62.4 | 18.1 | 20.3 | 21.3 |
| Cystic Fibrosis (%) | 0 | 1.7 | 0.4 | 0.5 | 0.6 | 0.2 | 0.3 | 0.6 | 0.2 | 0.2 |
| Neurosis (%) | 0.2 | 0.3 | 0 | 0 | 0 | 0 | 0.2 | 0.1 | 0 | 0 |
| Psychoses (%) | 1.3 | 5.5 | 1.9 | 1.8 | 1.1 | 1.3 | 6.7 | 1.7 | 1.5 | 8.3 |
| Eating Disorders (%) | 0 | 0.2 | 0.1 | 0.1 | 0 | 0 | 0.2 | 0 | 0 | 0.6 |
| Disrupt Childhd Disorders (%) | 0.2 | 0.1 | 0.1 | 0 | 0 | 0 | 0.2 | 0 | 0 | 0.2 |

| | | | | | | | | | |
|---|---|---|---|---|---|---|---|---|---|
| Substance Reltd Disorders (%) | 7 | 19.6 | 4.4 | 22 | 5.4 | 2.9 | 28.1 | 5.3 | 3.8 | 7.9 |
| Skin Cancer (%) | 4.9 | 3.8 | 5.8 | 5.8 | 19.5 | 5.6 | 7 | 21.4 | 11 | 9.1 |
| Congenital Heart Disease (%) | 0.8 | 0.4 | 1.5 | 1.6 | 0.4 | 0.5 | 0.9 | 1.1 | 1.5 | 1.4 |
| Periodontal Disease (%) | 0.8 | 0.7 | 0.6 | 0.4 | 1.1 | 0.7 | 0.4 | 0.4 | 0.4 | 0.6 |
| Chronic Fatigue Syndrome (%) | 0.6 | 0.6 | 0.6 | 0.9 | 0.4 | 0.2 | 1.6 | 1.1 | 0.6 | 0.6 |
| Fibromyalgia (%) | 3.9 | 3.8 | 1.2 | 2.7 | 3.1 | 0.7 | 14.9 | 1.2 | 2.2 | 2 |
| Parkinson Disease (%) | 1.3 | 0.9 | 1.6 | 2.8 | 2.3 | 2.7 | 2.1 | 1.9 | 1.7 | 15.4 |
| Hypercoaguable Syndome (%) | 1 | 1.1 | 0.8 | 1.2 | 1.3 | 1.2 | 1.7 | 0.9 | 1.4 | 1.4 |
| Post Partum BH Disorder (%) | 0 | 0.2 | 0 | 0 | 0 | 0 | 0 | 0 | 0 | 0 |
| Metabolic Syndrome (%) | 23.3 | 6.8 | 3.3 | 3.2 | 3.6 | 2.5 | 7 | 5.7 | 4.8 | 4.3 |
| Psyc Dis rltd Med Condtns (%) | 1.5 | 2.9 | 1.5 | 2.2 | 1.6 | 2 | 4.3 | 1.3 | 2.9 | 23.1 |

*Table 1: Summaries of disease distributions in different clusters. This table shows cluster summaries for age (median; years), male and female composition (%), and the proportion of people identified with different medical conditions (%; number members in cluster with the disease/ total number of members with the disease), for the nine clusters.*

5. Discussion:

In the present study we use the HAC approach to categorize clusters of patient with analogous multimorbidities in the Ohio population. The clusters identified were consistent and clinically applicable with relevant insights. We describe a rapid, data driven, scalable method to discover multimorbidity cohorts in the patient data. HAC on the Ohio patient health data to isolate multimorbidity patients revealed ten well defined clusters. An analysis of the most prevalent conditions in every cluster revealed broad groupings within each cluster (Table 2).

Our first cluster had 1,193 patients. It contained middle aged patients (median age 68.5 years) and had a slightly higher ratio of females (58%) compared to females (42%). The cluster was characterized by the highest incidence of oral cancer (87%) and metabolic syndrome (23%). Indeed, a 2015 meta-analysis revealed an association between metabolic syndrome and oral cancer[22]. Our second cluster had a higher ratio of younger females (55% females with median age of 62.2 years) and had the highest incidences of conditions like ovarian cancer (5.1%), cystic fibrosis (1.7%), pancreatic cancer (2.1%), hepatitis (6.9%) and post-partum birth disorders (0.2%). This was also our youngest cluster. Recent research has shown a linkage between ovarian cancers, pancreatic cancer and post-partum disorders further strengthening these causal findings[23–25].

| Cluster | Summary |
|---|---|
| 1 | Metabolic Syndrome group |
| 2 | Cancer group |
| 3 | Peptic Ulcer group |
| 4 | Kidney Stones group |
| 5 | Hyperlipidemia and stomach cancer group |
| 6 | Heart Failure and Ischemic Heart Disease group |
| 7 | Obesity related conditions group |
| 8 | Epilepsy group |
| 9 | Osteoarthritis group |
| 10 | Osteoporosis group |

*Table 2: A summary of clinical findings from each cluster.*

Cluster three was an older cluster (median age 75.7 years), with more males compared to females (58% males). This cluster contained the highest percentage of members with stomach conditions including peptic ulcers (82%), cholecystitis (99%), non-specific gastric dyspepsia (79%). Interestingly, we also found the highest incidence of lung cancer (23%) in this patient cohort. This association has previously been attributed to metastasis[26].

Cluster 4 was gender equal and contained a high proportion of older patients with (median age 76.4 years; 51 % females; 49% male) with 92% of the members with kidney stones. Other conditions also seen in this cluster included colorectal cancer (4%), hypertension (67%), multiple sclerosis (13%) and congenital heart disease (2%). Alexander and colleagues, in a retrospective study have previously shown the how chronic hypertension is strongly linked and indicative of calcium oxalate crystal formation hinting at the root cause of this association[27]. Cluster 5, contained 1139 patients and a very high proportion of patients with hyperlipidemia (90%) and stomach cancer (4%). Recent research has shown significant association between the two conditions[28,29].

Cluster 6 contained 1,494 patients, was more male skewed (57%), with a median age of 72.9 years. This cluster was predominantly dominated by heart diseases like heat failure (0.7%) and ischemic heart disease (0.4%).

Cluster 7 also had a high proportion of young females (67% females; median age 69.9 years) who reported a myriad of related disorders including chronic thyroid conditions (2.9%), endometrial cancer (3.3%), female infertility (0.2%), disrupted childhood disorder (0.2%) and obesity (24.8%). Similar multi-morbidity associations have extensively been studied in these patientis[30–32].

Cluster 8, was one of the older clusters (median age 77.5 years) with the highest proportion of males (59.3% males) with a majority of the members with epilepsy (78%). Epilepsy, one of the most prevalent neurological conditions with no known cure, is frequently seen in this subgroup[33,34]. Cluster 9 had more females (52%) with a median age of 79.2 years with a high proportion of patients with osteoarthritis (37%) and low vision blindness (52%). Cluster 10, our smallest cluster with only 507 members, was also our oldest with a median age of 81.1 years and a higher proportion of females (57%). This cluster consisted of diseases like osteoporosis (87%) and, rheumatoid arthritis (71%) and diverticular disease (64%). Osteoporosis and rheumatoid arthritis have often been linked to have similar mechanisms linked to interleukin-6 dysfunction [35–37] and this association further strengthens this hypothesis. This study has identified clinically relevant multimorbid clusters in a general population and can thus be used to profile patients in other geographies.